\documentclass{aa}  

\usepackage{graphicx,txfonts,natbib,twoopt}
 \usepackage[breaklinks=true]{hyperref} 
 \bibpunct{(}{)}{;}{a}{}{,}    
 \newcommandtwoopt{\citeads}[3][][]{\href{http://adsabs.harvard.edu/abs/#3}%
                                        {\citealp[#1][#2]{#3}}}
 \newcommandtwoopt{\citepads}[3][][]{\href{http://adsabs.harvard.edu/abs/#3}%
                                        {\citep[#1][#2]{#3}}}
 \newcommandtwoopt{\citetads}[3][][]{\href{http://adsabs.harvard.edu/abs/#3}%
                                        {\citet[#1][#2]{#3}}}
 \newcommandtwoopt{\citeyearads}[3][][]%
   {\href{http://adsabs.harvard.edu/abs/#3}{\citeyear[#1][#2]{#3}}}

\begin{document}

   \title{Rotational modulation of the linear polarimetric variability  of the cool dwarf TVLM 513--46546}
 \author{P. A. Miles-P\'aez
          \inst{1}\fnmsep\inst{2}
          \and
          M. R. Zapatero Osorio\inst{3}
          \and
           E. Pall\'e\inst{1}\fnmsep\inst{2}
          }

   \institute{Instituto de Astrof\'\i sica de Canarias, Calle V\'ia L\'actea s/n, 38205 La Laguna, Spain\\
              \email{pamp@iac.es, epalle@iac.es}
         \and
             Dpt. de Astrof\'isica, Univ. de La Laguna, Avda. Astrof\'isico Francisco S\'anchez s$/$n, 38206 La Laguna, Tenerife, Spain.\\
          \and
             Centro de Astrobiolog\'\i a (CSIC-INTA), Carretera de Ajalvir km 4, 28850 Torrej\'on de Ardoz, Madrid, Spain\\
             \email{mosorio@cab.inta-csic.es}
             }


  \abstract
   {}
   {We aimed to monitor the optical linear polarimetric signal of the magnetized, rapidly rotating M8.5 dwarf TVLM\,513$-$46546.} 
   {$R$- and $I$-band linear polarimetry images were collected with the ALFOSC instrument of the 2.56-m Nordic Optical Telescope on two consecutive nights covering about 0.5 and 4 rotation cycles in the $R$ and $I$ filters, respectively. We also obtained simultaneous intensity curves by means of differential photometry. The typical precision of the data is $\pm$0.46\%~($R$), $\pm$0.35\%~($I$) in the linear polarization degree and $\pm$9 mmag ($R$), $\pm$1.6 mmag ($I$) in the differential intensity curves.}
   {Strong and variable linear polarization is detected in the $R$ and $I$ filters, with values of maximum polarization ($p^{*}$ = 1.30\,$\pm$\,0.35\,\%) similar for both bands. The intensity and the polarimetric curves present a sinusoid-like pattern with a periodicity of $\sim$1.98 h, which we ascribe to structures in TVLM\,513$-$46's surface synchronized with rotation. We found that the peaks of the intensity and polarimetric curves occur with a phase difference of 0.18\,$\pm$\,0.01, and that the maximum of the linear polarization happens nearly half a period (0.59\,$\pm$\,0.03) after the radio pulse. We discussed different scenarios to account for the observed properties of the light curves.}
   {}


   \maketitle
%

\section{Introduction}\label{sec1}
The M8.5 dwarf TVLM 513--46546  (TVLM 513--46; \citeads{1993AJ....105.1045T}) is one of the most studied ultracool dwarfs in the literature. With a trigonometric distance of $10.76\,\pm\,0.03$ pc \citepads{2013ApJ...777...70F}, an effective temperature ($T_{\rm eff}$) of $2175\,\pm\,150$ K, a luminosity of log\,($L/L_\odot$) = $-3.59 \pm 0.02$ dex  \citepads{2004AJ....127.3516G}, and a lack of Li at 670.8 nm \citepads{1994ApJ...436..262M,2002AJ....124..519R}, the models of \citetads{2003A&A...402..701B} constrain the mass of TVLM 513--46 to the range $0.06-0.08$ M$_{\sun}$ and its radius to 0.1 R$_{\sun}$ for ages older than 0.5 Gyr. It shows H$\alpha$ in emission that changes moderately with time, which suggests some chromospheric activity \citepads{2002AJ....124..519R}. From its high projected rotation velocity ($v$\,sin\,$i\,=\,60\,\pm\,2$  km\,s$^{-1}$;  \citeads{2003ApJ...583..451M}) and its rotation period ($1.959574\,\pm\,0.000002$ h, \citeads{2014ApJ...788...23W}), the inclination of its rotation axis is found to be $i\,=\,74.5^{+10.2}_{-5.8}$ deg. The detection of radio emission reveals a multipolar magnetic field, with intensities as high as 3 kG \citepads{2006ApJ...653..690H,2008ApJ...673.1080B}. This magnetic field and the detection of  optical variability have led to a rich discussion on which physical scenario is the most likely to explain its variability: magnetic spots or dust clouds \citepads{2007ApJ...668L.163L,2008MNRAS.391L..88L,2008ApJ...673.1080B,2013ApJ...779..101H}.

Because the identification of this scenario is degenerate, additional observational parameters which add new information from this objects's atmosphere are necessary. In particular, the amount of linear polarization suffered by the light emitted/scattered toward a distant observer, and its variability, could be a very powerful discrimination tool. In this work, we present our $R$- and $I$-band monitoring of the intensity and linear polarimetric properties for TVLM 513--46 to try to shed more light on the physical processes responsible for the optical modulation.


\section{Observations}

\begin{figure}
   \begin{center}
    \includegraphics[width=0.32\textwidth]{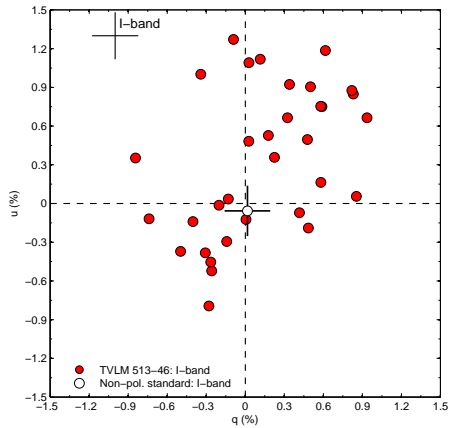}
     \caption{Stokes $q-u$ plane for the $I$-band measurements of TVLM\,513$-$46 (red circles) taken during the second night of observations. Instrumental polarization is shown with a white circle. The typical uncertainties in $q$ and $u$ are represented in the top-left.}
              \label{qu}
     \end{center}
    \end{figure}

\begin{figure}
   \begin{center}
    \includegraphics[width=0.4\textwidth]{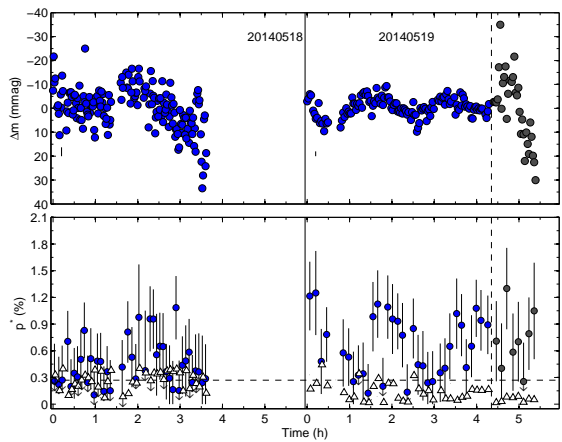}
     \caption{Differential intensity {\sl (top)} and debiased linear polarimetry {\sl (bottom)} curves of TVLM\,513$-$46 taken on 2013 May 18 (left) and 19 (right). The vertical dashed lines separate the $I$- (blue circles) and the $R$-band data (gray circles). TVLM\,513$-$46 is shown with solid dots. The reference star is depicted with white triangles and no error bars (for clarity) in the bottom panel.  The horizontal dashed line stands for the upper limit on instrumental polarization. Zero time corresponds to Julian dates of 2456431.5734 (2013 May 18) and 2456432.4922 (2013 May 19).}
              \label{Fig2_app}
     \end{center}
    \end{figure}

We collected $R$- ($\lambda_c$ = 631 nm) and $I$-band ($\lambda_c$ = 810 nm) linear polarimetry images of TVLM 513--46 using the Andaluc\'\i a Faint Object Spectrograph and Camera (ALFOSC) mounted on the 2.56-m Nordic Optical Telescope (NOT) on 2013 May 18 and 19. The target was monitored over $\sim$3.5 and $\sim$4.2 h ($I$-band) during the first and second nights, and over $\sim$1 h ($R$-band) during the second night. We thus covered $\sim4$ rotation cycles in the $I$ filter and half a rotation cycle in the $R$-band with a typical cadence of 6.2 and 9.6 min. Using these data, we also retrieved the $R$- and $I$-band intensity curves of TVLM 513--46 by means of differential photometry and by combining the polarimetric ordinary and extraordinary rays. The star 2MASS\,J15011008$+$2250069 (2.05 mag brighter in $I$ and located near our target) acted as the reference/comparison source. The detailed description of the observing strategy, instrumental configuration, and data reduction is provided in Sect.~\ref{sec2} (online material). The normalized Stokes parameters $q$ and $u$, the degree of linear polarization ($P$), and the polarization vibration angle ($\Theta$) were derived using the flux-ratio method and equations 1--4 from \citetads{2005ApJ...621..445Z}. The final data have typical uncertainties as follows: $\pm$0.46\%~($R$) and $\pm$0.35\%~($I$) in $P$, $\pm$9\degr~in $\Theta$, and $\pm$3.4 mmag ($I$, first night), $\pm$1.6 mmag ($I$, second night) and $\pm$9 mmag ($R$) in the differential intensity curves. Figure~\ref{qu} illustrates the $I$-band $q-u$ plane obtained using the second observing night measurements. We shall focus on these data because they have better quality by a factor of 2 than the photometry of the first night (see also Fig$.$~\ref{Fig2_app}). This allows us to study the details of the polarimetric light curve of TVLM\,513$-$46. The data of the first observing night are presented in Fig$.$~\ref{Fig2_app} and  Section A; they support our findings within the quoted error bars.  

%


%
\section{Light curves}\label{sec3}

Figure~\ref{Fig2_app} (top panel) illustrates the $R$- and $I$-band differential intensity curves of TVLM 513--46 as a function of time. The photometry exhibits sinusoidal variability, a Lomb-Scargle periodogram of our curves\footnote[1]{We used our own codes and the codes provided by the NASA exoplanet archive on: http://exoplanetarchive.ipac.caltech.edu/cgi-bin/Periodogram/nph-simpleupload} \citepads{1982ApJ...263..835S} shows a peak at $1.9\,\pm\,0.4$ h (Figure~\ref{figper}), which fully agrees with the period reported by other groups \citep[and references therein]{2014ApJ...788...23W} and confirms the stable periodic variability status of TVLM 513--46. This period likely reflects the rotational modulation of TVLM 513--46. The error bar of $\pm 0.4$ h was determined as the full width at half maximum of the 1.9-h peak of the periodogram. The $I$-band amplitudes of the differential intensity curves are measured at $\sim$4.9 mmag (first night) and $\sim$3.9 mmag (second night), which are intermediate between those given for filters of related wavelengths by \citet{2007ApJ...668L.163L} and \citetads{2008MNRAS.391L..88L}. Our measurements support the changing amplitudes of the observed sinusoidal variations with time. Because the periodicity is constant, this suggests that the intensity and/or size of the feature responsible for the variations is also changing (see also \citealt{2013ApJ...779..101H}).

The bottom panel of Fig.~\ref{Fig2_app} depicts the evolution of the $R$- and $I$-band linear polarimetry of TVLM 513--46. Because of the better signal-to-noise ratio of the second night photometry, the modulation of the polarimetric light curve becomes apparent on the right panel. Values between zero and strong linear polarization of $p^* \approx 1.30\% \pm 0.35\%$ are seen at the two wavelengths of our study. As a consistency check, we also derived the linear polarimetry photometry of the reference star in the same manner as for the target. These measurements are included in Fig.~\ref{Fig2_app} (bottom panel); they uniformly lie below the upper limit on the instrumental polarization, as expected for an unpolarized source. Because the reference star is located at a projected separation of $\sim$27\arcsec~from our dwarf, we concluded that the observed high linear polarization is intrinsic to TVLM 513--46 and is not due to an interloper cloud. Furthermore, the dwarf lies at a distance of 10.76\,$\pm$\,0.03 pc \citep{2013ApJ...777...70F} in a direction with negligible extinction by the interstellar medium. The solar vicinity ($\le $25 pc) shows very little linear polarization (less than 0.0004\%~per parsec, \citeads{1977A&AS...30..213P}). Only at distances $\gtrsim$70 pc, interstellar polarization  becomes non-negligible \citepads{2002A&A...394..675T}. 

Interestingly, the maximum polarization signal of TVLM 513--46 is similar for both the $R$ and $I$ filters, although we caution that the $R$-band observations are not simultaneous with the $I$-band and that the former data cover only a fraction of the rotation period. The Lomb-Scargle periodogram of the $I$-band linear polarimetry light curve shows a peak at $\sim$2-h (Figure~\ref{figper}) with a confidence of 83\% (see the Appendix). This agrees with the periodicity previously derived from the intensity curve and with the very precise rotation periods given by \citet{2013ApJ...779..101H} and \citet{2014ApJ...788...23W}. A similar Lomb-Scargle analysis of the polarimetric photometry of the reference star yielded no significant peaks at around 2 h. Our data indicate that TVLM\,513$-$46 is linearly polarized and that the intensity of the polarization changes with a periodicity compatible with the dwarf's rotation.

   \begin{figure}
   \begin{center}
    \includegraphics[width=0.35\textwidth]{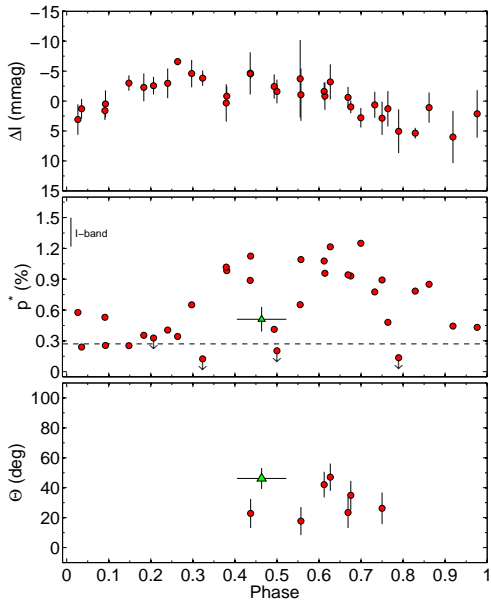}
     \caption{$I$-band differential intensity (top), linear polarimetry (middle), and polarization vibration angle (bottom) curves folded in phase using a periodicity of 1.9798 h. In the top panel, each data point stands for the average of four individual measurements and its associated dispersion. In the bottom panel, $\Theta$ is plotted for $P/\sigma_P \ge 2.7$. Two phases are presented for clarity. The green triangle stands for a $J$-band measurement taken from \citetads{2013A&A...556A.125M}}
              \label{Fig3}
     \end{center}
    \end{figure}

The intensity and linear polarimetry light curves of the M8.5 dwarf are folded in phase using a period of 1.9798 h in Fig.~\ref{Fig3} (and in Fig. \ref{Fig3_app} for data of both nights). Despite the short-time coverage of the $R$-band observations and their relatively large associated uncertainty, our data indicate that the intensity variations in the $R$ and $I$ wavelengths are correlated (Fig. \ref{Fig3_app}), as opposed to the anti-phased $g$ and $i$ light curves discussed by \citetads{2008MNRAS.391L..88L}. The previous $J$-band (1.2 $\mu$m) linear polarimetry measurement by our group \citepads{2013A&A...556A.125M} is also included. It was taken 3.7 months prior to the ALFOSC observations.  To improve the S/N ratio of the $I$-band intensity data and to equalize the time sampling of both light curves, we plot the average of four individual data points in the top panel. Within the observed scatter, the intensity and polarimetric light curves exhibit sinusoid-like variations, which we ascribe to large-scale structures in the dwarf's surface that move with rotation. There are two striking properties of the linear polarimetry light curve: one is that the maximum of the $I$-band linear polarization occurs after the peak of the intensity curve. By applying simple sinusoidal fits to the data, we derived that both peaks are phase shifted by $0.18\,\pm\,0.01$ or about 21 min. \citetads{2014ApJ...788...23W} reported that the radio emission bursts of TVLM\,513$-$46 are detected 0.41\,$\pm$\,0.02 of the period earlier than the peak of the intensity curve, which indicates the dipolar nature of the magnetic field and the inclination of the magnetic field with respect to the rotation axis \citepads{2009ApJ...695..310B}. Therefore, there is a delay of nearly half a rotation (0.59\,$\pm$\,0.03) between one burst in the radio emission of the M8.5 dwarf and its $I$-band polarization maximum. 

The highest $I$-band linear polarization degrees take place during phases $\sim$0.4 through $\sim$0.8, while the intensity light curve keeps decreasing toward the minimum brightness (top and middle panels of Fig.~\ref{Fig3}). The polarization vibration angle is shown folded in phase in the bottom panel of Fig.~\ref{Fig3}. With few exceptions, all $I$-band measurements are consistent with the mean value of $\Theta$\,=\,31\degr$\pm$10\degr, thus we do not resolve any particular pattern in the distribution of $\Theta$ given the precision of the data. Furthermore, the $I$- and $J$-band polarization angles are compatible at the 1-$\sigma$ with the quoted uncertainties, which suggests the little dependence of $\Theta$ on wavelength and/or the stability of the polarizing feature for at least $3.7$ months.

\section{Interpretation}\label{sec4}
The shape and characteristics of the polarimetric and intensity light curves of TVLM\,513$-$46, combined with other well-known properties of the M8.5 dwarf, may provide evidence of the physical origins of its linear polarization. Because of the coincidental periodicities between the two data sets, it is likely that the polarimetric variability is related to structures that spin with the dwarf's rotation. For example, a (cloudy or magnetic) spot on the dwarf's surface can change the intensity curve in a periodic manner, which will also be reflected in the polarization. Next, we discuss the following linear polarization mechanisms: Magnetic Zeeman splitting, scattering processes by electrons and dusty particles, and dichroic extinction. 

TVLM\,513$-$46 harbors a large scale, steady magnetic field, which is responsible for the observed radio emission \citepads{2006ApJ...653..690H,2008ApJ...673.1080B}.  \citetads{2002A&A...396L..35M} and \citetads{2004sf2a.conf..305M} studied 21 M1--L5 dwarfs located at distances of less than 32 pc, finding values of $P\lesssim0.2\%$ ($R$ and $I$ filters). We remark that some of the stars analyzed by these authors have magnetic fields with strengths of a few to several kG \citepads{2006Sci...311..633D,2007ApJ...656.1121R}, similar in intensity to TVLM\,513$-$46. Nevertheless, magnetic Zeeman broadening and splitting of individual lines due to species like FeH and CrH in very cool M and L dwarfs have been observed and modeled theoretically (e.g., \citealt{2006ApJ...644..497R,2013A&A...558A.120K}). This yields significant linear polarization as a function of the magnetic field strength at very specific wavelengths. When convolved with high rotation velocities and broad-band filters, the polarimetric signal is notoriously reduced but the net integration over broad bands may be different from zero, as pointed out by \citetads{2013A&A...558A.120K}. The $I$-band filter of our observations covers features due to Na\,{\sc i}, TiO, VO, CrH, FeH, H$_{\rm2}$O, ... in the wavelength interval 0.720--0.875\,$\mu$m, some of which are magnetically sensitive; therefore, we cannot discard a small contribution of this effect to the observed polarization. However, it may contribute with less than $P \approx 0.2\%$ as suggested by the observations of \citet{2002A&A...396L..35M} and \citetads{2004sf2a.conf..305M}. As we detect very high values of linear polarization spanning over 40\%~of the rotation period, the Zeeman broadening/splitting due to a strong magnetic field is not sufficient to account for all of the observed optical linear polarization of TVLM\,513$-$46. 

In addition, the distribution of our polarimetric measurements in the $q-u$ plane of Fig.~\ref{qu} hints at polarizing mechanisms related to scattering processes. Such a distribution is reproduced by \citetads{1986A&A...169..251C}, who modeled the polarimetric behavior of a globule of particles in orbit around a central unpolarized source using the Thompson scattering scenario (see their Figs. 3b and 5a). For an equator-on source and a cloud of particles located close to this region, the Stokes parameters describe a near circular pattern like that of Fig.~\ref{qu} (also Fig.~\ref{qu_app}). The $q-u$ distribution shape changes significantly as a function of the location of the cloud across the disk of the source and the angle between the magnetic and rotational poles. The strong magnetic field of TVLM\,513$-$46 reveals the presence of a surrounding envelope of low pressure electrons. Actually, the 100\%~circularly polarized radio pulses of the M8.5 dwarf are believed to be most likely caused by the electron-cyclotron maser instability \citep{2008ApJ...684..644H}, while the quiescent radio emission is probably due to gyrosynchrotron radiation. Our data are consistent with the $R$- and $I$-band linear polarization being originated by the Thompson scattering of an assembly of free electrons in a similar manner to what is observed in hot Be stars (e.g., \citealt{1978A&A....68..415B,1986A&A...169..251C}). Under this scenario, the electronic cloud must be located close to one radius of TVLM\,513$-$46 to produce polarimetric variability synchronized with rotation. \citet{2011A&A...525A..39Y} and \citet{2012A&A...539A.141K} determined that the electron density is of the order of 10$^5$ cm$^{-3}$ in TVLM\,513$-$46's magnetosphere, which is confined within two stellar radii according to \citetads{2011AJ....142..189J}.

Dust, which is present in primordial and debris disks, can also act as an efficient polarizer. Infrared flux excesses provide evidence of the presence of any of these disks surrounding central objects. TVLM\,513$-$46 does not appear to have clear flux excesses at 24 $\mu$m \citepads{2007ApJ...667..527G}, which thus discards the existence of warm primordial and debris disks. In addition, the M8.5 dwarf has efficiently depleted lithium from its atmosphere \citepads{2002AJ....124..519R}, which indicates a minimum age of a few hundred Myr (also see \citealt{2008ApJ...684..644H}). Dust is also present in the photospheres of dwarfs cooler than $T_{\rm eff}$\,=\,2700 K (e.g., \citeads{1996A&A...305L...1T}) as a result of the natural chemistry of low-temperature and high-gravity atmospheres that includes the condensation of refractory elements into liquid and solid particles. Under the scenario of large rotation velocities (that induce oblate shapes) and/or inhomogeneous distributions of dusty clouds, \citetads{2010ApJ...722L.142S} and \citetads{2011MNRAS.417.2874M} theoretically showed that the net linear polarization of dusty dwarfs by single scattering processes can be as high as a few per cent at optical and near-infrared wavelengths. \citet[and references therein]{2013A&A...556A.125M} provided  observational proof of linear polarization detections in ultracool dwarfs. TVLM\,513$-$46 has $T_{\rm eff}$ = 2025--2480 K  \citepads{2004AJ....127.3516G,2009ApJ...702..154S} and a rapid rotation (1.96 h and $v$\,sin\,$i$\,=\,60 km\,s$^{-1}$); condensates (dust particles) are thus likely populating the upper (and coolest) atmospheric layers. If these particles are unevenly distributed across the atmosphere and/or aligned with the intense magnetic field, some linear polarimetry signal would be expected. 

\citetads{2011ApJ...741...59D} modeled the near-infrared polarization of a rotating dwarf with a 20\degr$\times$20\degr~dusty spot on its equator (Rayleigh scattering). They found a varying polarization degree and vibration angle with amplitudes of $\sim$0.6\%~and $\sim$40\degr~($\lambda=1.11\,\mu$m) as the cloudy spot moves across the object's visible disk with respect to the observer. As illustrated in Fig.~5 by \citetads{2011ApJ...741...59D}, the polarization degree has two maxima separated by a phase difference of 0.17 and a deep minimum in-between just when the cloud lies along the line of sight. There is a hint of this feature at phase 0.5 in the middle panel of Fig.~\ref{Fig3}. Regarding $\Theta$, the model predicts a sudden change in the angle within a phase difference of 0.055, which translates into 6.6 min in the case of the M8.5 dwarf. Our data do not have enough temporal sampling to test these predictions in $p^{*}$ and $\Theta$. \citetads{2011ApJ...741...59D} also calculated the intensity curve caused by the equatorial dusty spot. It shows a mild peak at the same phase as the deep minimum between the two polarimetric maxima. Our TVLM\,513$-$46 data do not recreate this signature since the maximum brightness of the $I$-band differential light curve and the deep minimum between the two maxima of the polarization curve are separated by $\sim$0.15 of the rotation period. This phase difference might be explained by changing the geometry in which the grains are illuminated. 

Finally, dichroic extinction by magnetically aligned non-spherical grains may lead to significant polarization \citepads{1951ApJ...114..206D}; this mechanism is believed to cause the interstellar polarization observed toward many stars in our Galaxy and has been used to map out the Galactic magnetic field \citepads{1970MmRAS..74..139M}. Under this scenario, the linear polarization of TVLM\,513$-$46 would lie parallel to the magnetic field direction, which is $31\,\pm\,10^{\circ}$. The periodic and sinusoid-like variation of $p^*$ would be caused by a cloud of aligned non-spherical grains rotating with the dwarf. The impact of convective motions and rapid rotation on such dusty clouds is not clear. Models of dichroic polarization particularized for ultra-cool dwarfs are required to explore this scenario further.

From our discussion it is clear that, with the present data, it is not posible to solve unambiguously the origin of the observed optical linear polarization. To this end, follow-up observations taken at shorter and longer wavelengths and with a higher cadence would also be beneficial: On the one hand, the intensity of the dichroic polarization has a strong dependence on wavelength, whereas the polarization vibration angle keeps constant \citepads{1963AJ.....68..185T,1974AJ.....79..565C}, and on the other hand, linear polarization due to other mechanisms shows a moderate variation with wavelength. Additionally, polarimetric observations combined with other techniques to map out the presence of dusty clouds, such as the tomography of the surfaces \citepads{2014Natur.505..654C} could be valuable tools to determine the geometry, distribution, and other properties of the clouds in the atmospheres of ultracool dwarfs, especially for those that do not harbor strong magnetic fields such as TVLM\,513$-$46.

\begin{acknowledgements}
This work was based on observations made with the Nordic Optical Telescope operated on the island of La Palma by the Nordic Optical Telescope Scientific Association (NOTSA) in the Spanish Observatorio del Roque de los Muchachos of the Instituto de Astrof\'\i sica de Canarias. This work is partly financed by the Spanish Ministry of Economics and Competitiveness through projects AYA2012-39612-C03-02 and AYA2011-30147-C03-03.
\end{acknowledgements}
%
%


\bibliographystyle{aa} 
\bibliography{biblio.bib} 


\Online
\begin{appendix} 
\section{Observations and data reduction}\label{sec2}

We collected linear polarimetry images of TVLM 513--46 using the Andaluc\'\i a Faint Object Spectrograph and Camera (ALFOSC) mounted on the 2.56-m Nordic Optical Telescope (NOT) on 2013 May 18 and 19. ALFOSC has a 2048$\times$2048 E2V detector with a pixel size of 0\farcs19. The target was monitored in the $I$-band ($\lambda_c$ = 810 nm) over $\sim3.5$ and $\sim4.2$ h during the first and second nights, and in the $R$-band  ($\lambda_c$ = 631 nm) over $\sim1$ h during the second night. We thus covered $\sim4$ rotation cycles in the $I$-band and half a cycle in the $R$-band. Sky conditions were clear and seeing varied in the interval 1\farcs1--1\farcs5.  Observations were carried out at air masses ranging from 1.05 through 2.34.

\begin{figure}
   \begin{center}
    \includegraphics[width=0.49\textwidth]{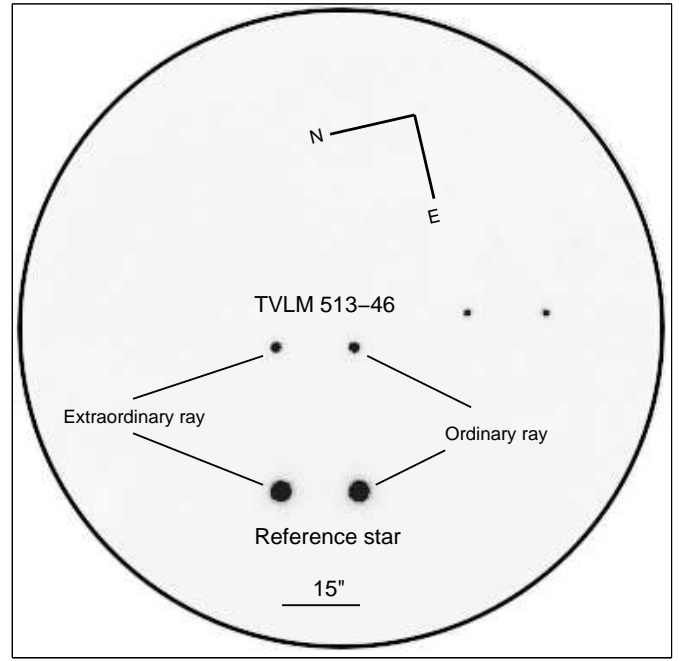}
     \caption{ALFOSC linear polarimetric $I$-band image of TVLM\,513$-$46. The circle indicates the unvignetted area.}
              \label{Fig1}
     \end{center}
    \end{figure}

The linear polarimetry observing mode of ALFOSC consists of a half-wave plate and a calcite block, which provides simultaneous images of the ordinary and the extraordinary beams separated by 15\arcsec. The total unvignetted field of view is 140\arcsec~in diameter, which we rotated by 103\degr~East of North to align TVLM 513$-$46 and a nearby, bright reference star along the $y$-axis of the detector, as it is indicated in Fig.~\ref{Fig1}. Individual exposure times were 50 s ($I$, first night) and 100 s ($R$ and $I$, second night) per position of the half wave plate (0\degr, 22\fdg5, 45\degr, and 67\fdg5). On May 18, polarimetric images were taken at two nod positions on the detector separated vertically by 10\arcsec~for a proper sky subtraction. This nodding pattern was not applied on the following night since TVLM 513$-$46 is detected well above the sky contribution with signal-to-noise (S/N) ratio of $\sim$220 and $\sim$300 ($I$-band) in 50-s and 100-s individual integrations, respectively (S/N is measured at the peak of the energy distribution of the source images). One polarimetric cycle was completed every 6.2 min ($I$, first night) and 9.6 min ($R$ and $I$, second night), including overheads. This allowed us to sample one rotation of TVLM 513$-$46 using a minimum of 12 and a maximum of 19 linear polarimetry measurements. Raw images were bias subtracted and flat-field corrected using data acquired during dawn and dusk and routines within IRAF\footnote[1]{IRAF (Image Reduction and Analysis Facility) is distributed by the National Optical Astronomy Observatories, which are operated by the Association of Universities for Research in Astronomy, Inc., under cooperative agreement with the National Science Foundation.}. We also observed non-polarized stars (GJ 838.4 and WD 1615--154, \citealt{2007ASPC..364..503F}) and one polarized source (Hilter 997, \citealt{1992ApJ...386..562W}) by employing the same instrumental configuration and on the same observing dates as the target. This allowed us to check for the efficiency of the instrument and to set an upper limit on the instrumental polarization.

%

   \begin{figure}
   \begin{center}
    \includegraphics[width=0.49\textwidth]{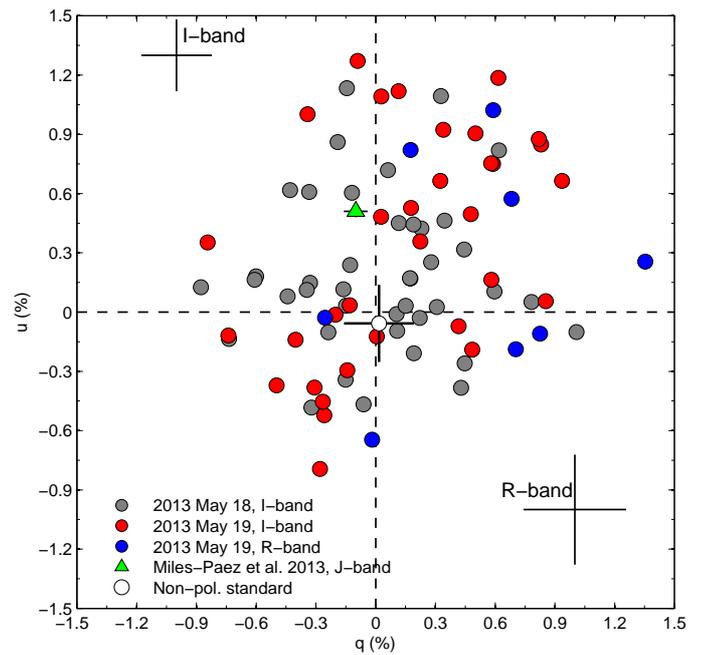}
     \caption{Stokes $q-u$ plane for the $R$- and $I$-band measurements of TVLM\,513$-$46) taken during first and second night of observations. Instrumental polarization is shown with a white circle. The typical uncertainties in $q$ and $u$ are indicated.}
              \label{qu_app}
     \end{center}
    \end{figure}
   \begin{figure}
   \begin{center}
    \includegraphics[width=0.49\textwidth]{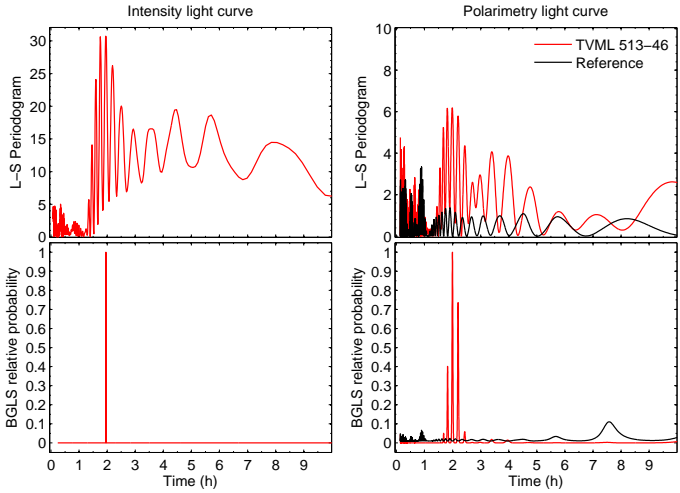}
     \caption{{\it Top:} Lomb-Scargle periodograms of the intensity (left) and polarimetric (right) light curves of TVLM 513--46 (red) and the reference star (black). {\it Bottom:} Relative probabilities of the explored periods using a Bayesian formalism for the generalized Lomb-Scargle periodogram (see text).}
              \label{figper}
     \end{center}
    \end{figure}

   \begin{figure}
   \begin{center}
    \includegraphics[width=0.45\textwidth]{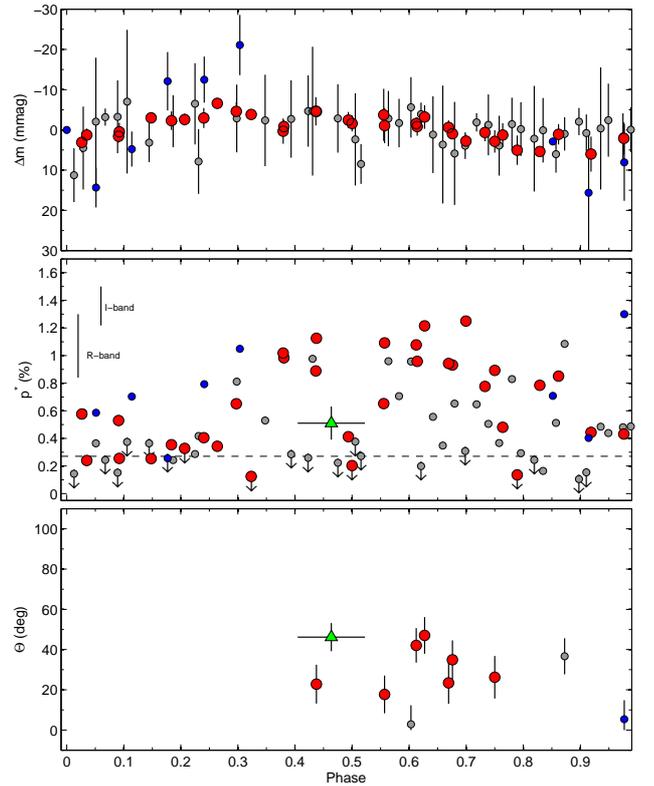}
     \caption{$R$- and $I$-band differential intensity (top), linear polarimetry (middle), and polarization vibration angle (bottom) curves folded in phase using a periodicity of 1.9798 h. Symbols as in Fig.~\ref{qu_app}. In the top panel, each data point stands for the average of four individual measurements and its associated dispersion. In the bottom panel, $\Theta$ is plotted for $P/\sigma_P \ge 2.7$. Two phases are presented for clarity.}
              \label{Fig3_app}
     \end{center}
    \end{figure}

We derived the normalized Stokes parameters $q$ and $u$, the degree of linear polarization ($P$), and the polarization vibration angle ($\Theta$) using the flux-ratio method and equations 1--4 from \citetads{2005ApJ...621..445Z}. The Stokes parameters and the angle $\Theta$ were properly corrected for the rotation of the field of view. Fluxes of each ordinary and extraordinary images were extracted by defining circular apertures ranging between 0.5 and 6 times the size of the full-width-at-half-maximum (FWHM) and using the PHOT package within IRAF. Sky annuli had inner radii ranging from 3.5\,$\times$ to 6\,$\times$\,FWHM and widths of 1.5\,$\times$\,FWHM. The detailed procedure is described in \citetads{2013A&A...556A.125M}. Our final $q$, $u$, $P$, and $\Theta$ values result from averaging the aperture photometry in the interval 2--4\,$\times$\,FWHM; the errors associated with $q$ and $u$ are defined as the standard deviations of the measurements in the selected range of apertures, the errors in $P$ are computed as the quadratic sum of the $q$ and $u$ quoted uncertainties plus the uncertainty introduced by a possible instrumental linear polarization (see below), and errors in $\Theta$ are determined following the equations given in \citetads{2013A&A...556A.125M} and \citetads{1974ApJ...194..249W}. We caution that the expression for the $\Theta$ uncertainty is valid for $P/\sigma_{P}\ge3$, where $\sigma_P$ is the error in linear polarimetry. Our linear polarimetric data of TVLM\,513$-$46 have a typical uncertainty of $\pm$0.46\%~($R$) and $\pm$0.35\%~($I$) in $P$, and $\pm$9\degr~in $\Theta$. 

Using the observations of the zero-polarized standard stars, we checked that the instrumental polarization lies below 0.31\%~and 0.27\%~in the $R$- and $I$-bands. The polarized standard star allowed us to determine the zero point correction for the polarization vibration angle to be $\Theta_{\circ, \,R}$\,=\,0\fdg8$\pm$2\fdg8 and $\Theta_{\circ,\,I}$\,=\,1\fdg3$\pm$3\fdg0, which agrees with the values tabulated in the ALFOSC  manual\footnote[2]{http://www.not.iac.es/instruments/alfosc/polarimetry/index.html}. From now on, we shall use the debiased linear polarization degree, $p^{*}$, defined as
\begin{equation}
p^{*}=\sqrt{P^{2}-\sigma_{P}^{2}},
\end{equation}
which accounts for an overestimation of the polarization signal at low S/N or small values of $P/\sigma_P$ \citepads{1974ApJ...194..249W}. We set $p^{*} = P$ if $\sigma_{P} \ge P$. In Figs$.$~\ref{Fig2_app}, \ref{Fig3}, and \ref{Fig3_app}, these polarimetric data are displayed with an associated arrow indicating that the true polarization index likely lies between 0\%~and the symbol position. We note that the dispersion of consecutive $\Theta$ measurements is of the order of the uncertainty in the polarization vibration angle when $P/\sigma_{P}>2.7$, which is close to the regime where the calculation of $\Theta$ has statistical significance \citepads{1974ApJ...194..249W}. This provides support to our method for deriving the error bars associated with $q$, $u$, and $P$ or $p^{*}$. The individual Stokes parameters values for the $R$ and $I$ filters are shown in the $q-u$ plane of Fig.~\ref{qu_app}. 
   
In addition to the linear polarimetric evolution, we retrieved the $R$- and $I$-band intensity curves of TVLM 513--46 by means of differential photometry. The star 2MASS\,J15011008$+$2250069 (2.05 mag brighter in $I$ and located near our target) acted as the reference/comparison source. This is the only bright star in the ALFOSC field of view that was usable for the differential photometry technique (Fig. A.1). Because of its lower luminosity, the error bars in the differential intensity curves are dominated by the photon noise of TVLM 513--46. The typically 1.0\,$\times$\,FWHM-aperture fluxes of the ordinary and extraordinary images of TVLM 513--46 and its reference star were extracted with the VAPHOT package \citepads{2001phot.work...85D}, which works under the IRAF environment and is optimized for differential light curves. To build the intensity curves, ordinary and extraordinary fluxes were combined per individual frame, which thus provides a time sampling four times higher than the cadence of the linear polarimetry data. We estimated the uncertainties in the differential light curves to be $\pm$3.4 mmag ($I$-band) for the first-night photometry, and $\pm$1.6 mmag ($I$) and $\pm$9 mmag ($R$) for the second night data. We checked that the measured linear polarization degree and the differential intensity light curves do not correlate with airmass.

The Lomb-Scargle periodograms of the $I$-band intensity and polarimetric light curves of TVLM\,513$-$46 computed following \citetads{1976Ap&SS..39..447L} and \citetads{1982ApJ...263..835S} are illustrated in the top panels of Figure \ref{figper}. The periodogram of the polarimetric light curve of the reference star is also included in the Figure. We explored frequencies between $\omega=2\pi/T$ and $\omega=2N_{\rm o}/T$ with a spacing of about $1/T$ \citepads{1986ApJ...302..757H}, where $N_{\rm o}$ is the number of data points of the light curves (296 for the intensity curve, and 74 for the polarimetric curve), and $T$ is the total time coverage of the data (26.4 h). We thus surveyed periods in the range ~0.3--26.4 h using 433 and 88 independent frequencies for the intensity and polarimetric data, respectively. The confidence levels of the ~2-h peaks were estimated using the false-alarm-probability (1--FAP) at 99\%~(intensity light curve) and 83\%~(polarimetry). The bottom panels of Figure \ref{figper} depict the relative probabilities according to the Bayesian formalism described in \citetads{2015A&A...573A.101M}.


\end{appendix}

\end{document}